\documentclass[aps,epsfig,manuscript,12pt,superscriptaddress,preprintnumbers]{revtex4}

\usepackage{amssymb}
\usepackage{amssymb}
\usepackage{epsfig,graphicx,times}
\usepackage{enumerate}
\usepackage{dcolumn}
\usepackage{color}
\usepackage{bm}

\newcommand{\beq}{\begin{equation}}
\newcommand{\eeq}{\end{equation}}

\newcommand{\bea}{\begin{eqnarray}}
\newcommand{\eea}{\end{eqnarray}}

\begin{document}

\title{Theory of quantum energy transfer in spin chains: From superexchange to ballistic motion}
\author{Claire X. Yu}
\affiliation{Chemical Physics Group, Department of Chemistry and Center for Quantum
Information and Quantum Control, University of Toronto, 80 St. George
street, Toronto, Ontario, M5S 3H6, Canada}

\author{Lian-Ao Wu}
\affiliation{Department of Theoretical Physics and History of Science, The Basque Country
University (EHU/UPV) , 48080 Bilbao and IKERBASQUE -- Basque Foundation for Science, 48011,
Bilbao, Spain}

\author{Dvira Segal}
\affiliation{Chemical Physics Group, Department of Chemistry and Center for Quantum
Information and Quantum Control, University of Toronto, 80 St. George
street, Toronto, Ontario, M5S 3H6, Canada}

\date{\today}

\begin{abstract}
Quantum energy transfer in a chain of two-level (spin) units,
connected at its ends to two thermal reservoirs, is analyzed in two
limits: (i) In the off-resonance regime, when the characteristic
subsystem excitation energy gaps are larger than the reservoirs
frequencies, or the baths temperatures are low. (ii) In the
resonance regime, when the chain excitation gaps match populated
bath modes. In the latter case the model is studied using a master
equation approach, showing that the dynamics is ballistic for the
particular chain model explored. In the former case we analytically
study the system dynamics utilizing the recently developed
Energy-Transfer Born-Oppenheimer formalism [Phys. Rev. E {\bf 83},
051114 (2011)], demonstrating that energy transfers across the chain
in a superexchange (bridge assisted tunneling) mechanism, with the
energy current decreasing exponentially with distance. This behavior
is insensitive to the chain details. Since at low temperatures the
excitation spectrum of molecular systems can be truncated to
resemble a spin chain model, we argue that the superexchange
behavior obtained here should be observed in widespread systems
satisfying the off-resonance condition.
\end{abstract}


\maketitle

\section{Introduction}


The scaling of the energy current with system size is of interest
for developing applications in energy conversion \cite{Rachel},
molecular electronics \cite{MolEl}, and reaction dynamics
\cite{Rdyn}. In the context of biological macromolecules,
understanding pathways and efficiency of heat flow is important for
controlling signal transmission and functionality in biomolecules
\cite{Leitnerbook}. In nanoscale electric devices significant power
dissipation may lead to the system disintegration. Designing
efficient routes for energy transfer, away from the conducting
object, is essential for a stable operation \cite{Pop}.

Resolving the size effect of heat conduction in molecular chains has
been the subject of recent experiments \cite{Schwarzer, Segalman}.
For example, Wang et al. \cite{DlottE} has recently studied the
kinetics of heat transfer from a metal substrate to self-assembled
hydrocarbon monolayers of increasing sizes \cite{DlottRev},
concluding that heat travels ballistically along the chain
\cite{SegalHeat}. Vibrational energy transport in peptide helices
was measured by employing vibrational probes as local thermometers
at various distances from a heat source \cite{Hamm}. For this protein system
it was concluded that heat propagates in a diffusive-like process.

\begin{figure}
\hspace{-2mm} \vspace{-50mm} 
\epsfig{file=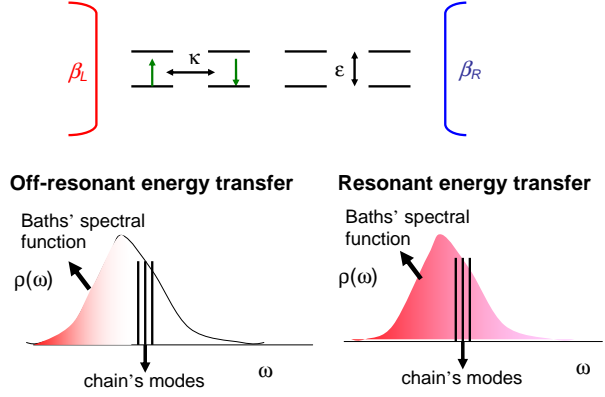, width=9cm} \caption{Top: Chain of two-level
particles or spin-1/2 objects coupled at the edges to heat baths maintained at
different temperatures. Bottom: In the off-resonance regime, applicable, e.g.,
at low temperatures,
the chain's modes, relevant for the
transport process, match unpopulated bath modes. In the opposite
resonance regime the chain's modes overlap with populated bath modes.
The coloring of the baths' spectral function represents  thermal population,
where darker color reflects larger occupation.
The off-resonant model could be also realized
considering reservoirs whose
spectral functions have a low cutoff, below the chain's characteristic excitation gaps.}
\label{FigS}
\end{figure}

Considering {\it electron transfer} across a 1-dimensional (1D)
conductor (bridge) connected at the edges to electron reservoirs,
multitude of theoretical, numerical, and experimental studies have
demonstrated that in the resonance limit, applicable for ohmic
reservoirs at high temperatures, (quasi) 1D chains conduct electrons
anywhere between a
ballistic to a diffusive manner, depending on the details of
the internal interactions. In contrast, in the off-resonance regime,
when the electronic levels of the bridge lie high, above the populated states of the
reservoirs (the donor and acceptor states in a donor-bridge acceptor
complex), a deep tunneling mechanism should take off \cite{NitzanRev}.

In this paper we focus on the analogous problem of {\it energy
transfer} in molecular chains connected at the two ends to thermal
reservoirs (baths), maintained at distinct temperatures, realized by
solids, metals, nanoparticles \cite{HammNP} or large molecular
complexes \cite{Schwarzer}. As a simple model for the central object
we consider a chain of several two-level units (spin 1/2
particles). The units are coupled through nearest-neighbor
coupling terms, assumed to be weak compared to the on-site
energies. Similarly to the electronic case, we expect that distinct transport
mechanisms will dominate at different parameter regimes.
For a schematic representation of the chain model and the relevant excitation spectra
see Fig. \ref{FigS}. 

Before proceeding, we carefully clarify our terminology: "Resonant"
regime refers here to the case where the modes of the chain,
responsible for the heat transfer dynamics, lie in resonance with
the occupied baths' modes. In contrast, "Off-Resonance" conduction
refers to the case where the occupied modes of the thermal
reservoirs' are low, below the typical excitation frequencies of the
enclosed object. Overall, there is always a conservation of energy
in our system, transferred between the two reservoirs (donor and
acceptor).
Thus, irrespective of the bridge energetics, we always
consider here a "resonance energy transfer" (RET) process
\cite{Scholes,EET}, in the sense that there are no energy loss
mechanisms, e.g., it is a non-radiative process.

In the resonant regime numerous theoretical and computational
studies have demonstrated that energy transfer between two
reservoirs, mediated by the excitation of the interlocated object,
may follow a ballistic $J\propto N^0$, ohmic $J\propto N^{-1}$, (or
somewhere in between, $J\propto N^{\alpha-1}$, $\alpha>0$)
mechanism. This was done in the context of vibrational energy
transfer \cite{Dhar} and thermal transport in spin chain
\cite{Brenig,Prosen,ProsenM}, for both classical and quantum
systems, using, e.g., molecular dynamics simulations \cite{Dhar},
the quantum master equation method \cite{Gemmer,ProsenM,Mona}, the
Green-Kubo formula \cite{Saito,MichelKubo}, and the density matrix
renormalization group method \cite{DMRG}. In the off-resonant regime
simulations on {\it purely harmonic} systems indicated on a
tunneling dynamics of heat transfer \cite{SegalHeat}. In the
presence of interactions, off-resonant quantum heat transfer
dynamics has been treated by adopting e.g., the complex machinery of
the Green's function approach \cite{Green1,Green2}, or by using
mixed quantum-classical simulations \cite{Stock}. Typically, such
methods only provide numerical results, hindering direct picture of
the microscopic processes involved.
Responding to this challenge,  we have recently developed a simple
analytic method for describing energy transfer in {\it nonlinear
systems} in the off-resonant regime \cite{BOheat}. This method, an
extension of the Born-Oppenheimer (BO) principle \cite{BO} to energy
transfer problems, can treat general subsystems (impurity, chains)
with intrinsic anharmonicities, as well as cases where the subsystem
is nonlinearly coupled to the reservoirs. The outcome of the method
is a Landauer type expression, incorporating nonlinear interactions
\cite{Land}, allowing for the identification of different scattering
processes \cite{BOheat}.
In what follows we refer to this method as the Energy-Transfer
Born-Oppenheimer (ETBO) scheme.

In this paper we aim in deriving scaling laws for the behavior of
the energy current with size for 1D molecular systems, primarily
focusing on the off-resonant limit \cite{comment}. For simplicity,
we consider the isotropic XY spin chain and its variants as a
prototype for a homogeneous and linear molecular chain
\cite{commXY}. This is a relevant physical description since at low
temperatures or in the off-resonance limit the energy spectra of the
interlocated system can be truncated, as transport predominantly
occurs through the lowest excitation states.
Considering a spin chain between two thermal reservoirs, we study
the energy transfer behavior in two different limits: (i) We assume
an off-resonance scenario, and obtain the energy current adopting
the ETBO approach. In this case we demonstrate that the energy
current decays {\it exponentially} with size, a footprint of the
tunneling mechanism. (ii) In a resonant situation we utilize a
standard master equation approach and show that the isotropic XY
chain behaves as a ballistic conductor, providing a fixed current
for different sizes. While we present our study in the context of
steady state heat transfer, the results are also useful for
interpreting energy transfer rates in donor-bride-acceptor complexes
\cite{Nitzan}

The tunneling behavior of the energy current resolved in the
off-resonance regime \cite{Speiser, Scholes,EET} resembles the
McConnell superexchange result \cite{superEx}, observed in electron
transport experiments in numerous systems, including monolayers
\cite{Mono}, proteins \cite{Protein}, and DNA \cite{DNA}. Since the
thermal superexchange result does not depend on the details of the
chain model, we expect it to show up in different physical systems
at low temperatures, including molecular wires, spin chains, and
biomolecules.

Our study here is presented in the context of thermal energy
transfer. However, the analysis and results are valid for describing
general {\it excitation energy transfer} problems (vibrational,
electronic), in bulk-molecule-bulk junctions and
 donor-bridge-acceptor systems \cite{Speiser,Scholes,EET}. Experiments on
$\sigma$-bond or $\pi$-bond bridges, connected to donor and acceptor chromophores
reported on excitation transfer rates which are exponentially
decreasing with size \cite{EETexp,EETratner}. This behavior is rigorously
recovered here.

The structure of this paper is as follows. In Section II we describe
the spin chain model, serving as a prototype for studying energy transfer and thermal conduction
in linear chains. In Section III we study the
off-resonant case using the ETBO method.
Perturbative analytic results are supported by
numerical simulations. Section IV treats the resonant limit,
adopting a master equation approach. Section V concludes.


\begin{figure}
{\hbox{\epsfxsize=90mm\epsffile{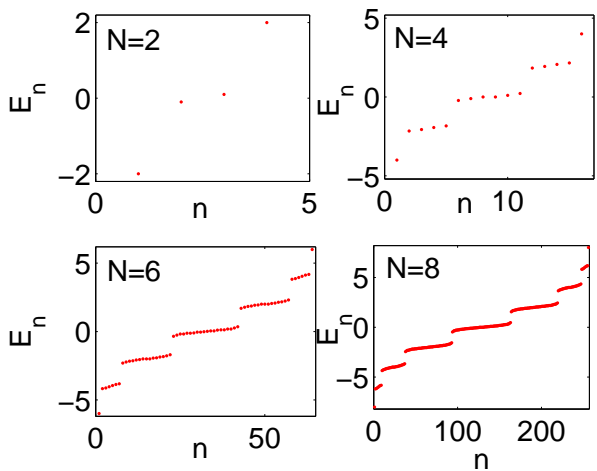}}} \caption{Energy spectrum
of the isolated isotropic XY spin chain with $N=2$, $N=4$, $N=6$, and $N=8$ units.
Other parameters are $\epsilon=2$ and $\kappa=0.1$. The different
manifolds include different numbers of excitations on the chain,
from zero up to $N$. } \label{Fig4}
\end{figure}

\begin{figure}
{\hbox{\epsfxsize=90mm\epsffile{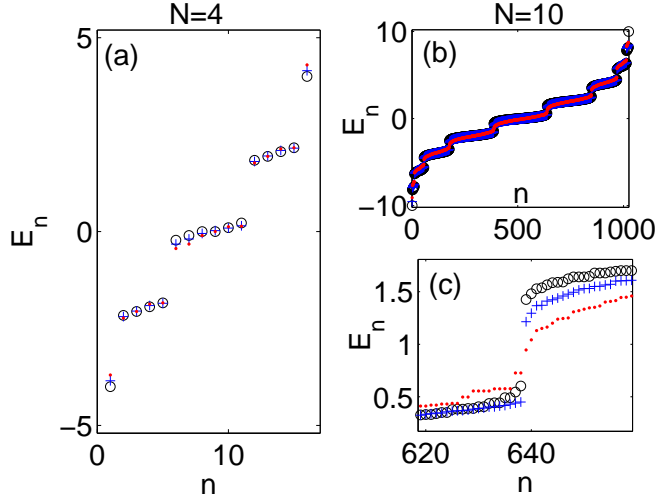}}} \caption{Energy spectrum
of the isolated anisotropic Heisenberg spin chain with $N=4$ (a) and $N=10$
(b). Panel (c) zooms on a small portion of the spectrum for $N=10$, manifesting
the gap closure with increasing $\delta$. Other parameters are $\epsilon=2$,
$\kappa=0.1$ and $\delta=0$ ($\circ$), $\delta=1$ ($+$) and
$\delta=2$ (dotted).} \label{Fig4a}
\end{figure}


\section{Model}

Consider a small subsystem, representing e.g., a molecule, placed in
between two thermal reservoirs (e.g., solids, large complexes) maintained each at a
fixed temperature $T_{\nu}=\beta_{\nu}^{-1}$ ($\nu=L,R$).
The total Hamiltonian is given by
\bea H=H_{S}+H_{L}+H_{R}+V_{L}+V_{R}. \label{eq:H} \eea
$H_{S}$ is the Hamiltonian of the subsystem and $H_{\nu}$ stands for
the $\nu$ heat bath. $V_{\nu}$ couples the subsystem and
the $\nu$ reservoir.
%
In particular, in what follows we focus on the thermal transport
properties of the isotropic XY spin-chain \cite{commXY}
\bea H_S=\frac{\epsilon}{2} \sum_{j=1}^N
\sigma_j^z+\frac{\kappa}{2}\sum_{j=1}^{N-1}\left(
\sigma_j^x\sigma_{j+1}^x+ \sigma_j^y\sigma_{j+1}^y \right).
\label{eq:Hspin} \eea
Here $\sigma_j^{x,y,z}$ are the Pauli matrices for the $j$ spin. The
first term describes the onsite spectrum at each site, a two level
system with a spacing $\epsilon$. The second term provides the
hopping interaction between neighboring sites with an exchange
strength $\kappa$. We generally assumes that $\kappa<\epsilon$,
allowing for a meaningful description of the chain in terms of its
subunits. The chain is coupled to two independent thermal reservoirs
at sites $1$ and $N$. Our derivation below does not make any
assumption regarding the structure of these baths. For example, they
may each contain a collection of independent harmonic oscillators
\bea H_{\nu}=\sum_{j\in \nu} \omega_{j}b_{\nu,j}^{\dagger}b_{\nu,j}.
\label{eq:Hbath} \eea
One may also consider fermionic reservoirs, a source of electronic
excitations. System-bath interactions are assumed to take the
following form,
\bea V_{L}&=&S_1 B_{L}; \,\,\,\, S_1=\lambda_L\sigma_1^x
\nonumber\\
V_{R}&=&S_N B_{R}; \,\,\,\, S_N=\lambda_R\sigma_N^x, \label{eq:HV}
\eea
where $\lambda_{\nu}$ parametrizes the system-bath coupling
strength, assumed to be a real number, $S_1$ ($S_N$) are subsystem
operators coupled to the left (right) reservoirs,  $B_L$ ($B_R$) are
$L$ ($R$) bath operators, which need not be specified at this stage.
This form assumes that the interaction with the reservoirs can
generate (absorb) an excitation at the leftmost or rightmost sites
of the chain.

Using the Jordan-Wigner transformation \cite{Jordan}, the (isolated)
isotropic XY chain can be reduced to describe spinless free
fermions. However, our analysis is still not trivial since the chain
is coupled, potentially in a nonlinear way, to thermal reservoirs
which are not necessarily modeled by isotropic XY chains by
themselves. Thus, the total Hamiltonian cannot be transformed into a
collection of noninteracting fermions, and the model is generally
non-integrable. Furthermore, the ETBO analysis can be carried out
for studying transport through the 1D anisotropic Heisenberg model
\cite{commXY},
\bea H_S^{}=\frac{\epsilon}{2} \sum_{j=1}^N
\sigma_j^z+\frac{\kappa}{2}\sum_{j=1}^{N-1}\left(
\sigma_j^x\sigma_{j+1}^x+ \sigma_j^y\sigma_{j+1}^y \right)
+\frac{\kappa\delta}{2}\sigma_j^z\sigma_{j+1}^z.
\label{eq:HspinHeis} \eea
The spectrum of $H_S$ with $\delta=0$ is exemplified in Fig.
\ref{Fig4} for several sizes. We note that for $\epsilon\gg\kappa$
subsystem energies are grouped into manifolds, each including
eigenstates with a particular number of excitations: At the bottom
of the spectrum lies a zero excitation state with an energy of
$E_0\sim (-\frac{N}{2}\epsilon)$. Above, we identify states
including a single excitation on the chain, with their energy
centered around $E_0+\epsilon$. The next manifold includes the
two-excitation states, and so on and so forth. For example, for
$N=4$ there are five manifolds with zero (bottom) to four (top)
excitations residing on the chain. Note that given the form of the
system-bath interaction operator [Eq. (\ref{eq:HV})], the thermal
baths, the excitation resources, can only translate the subsystem
between (neighboring) manifolds, adding or absorbing a single
excitation at a time. The reservoirs {\it cannot} directly drive
transitions {\it within} each manifold. Since for
$\kappa\ll\epsilon$ the typical gap between manifolds is $\sim
\epsilon$, this energy scale is identified as the characteristic
frequency of the subsystem, controlling the transport properties of
the model.

In Fig. \ref{Fig4a} we show that this picture is retained when the
exchange anisotropy parameter $\delta$ is relatively small,  for
short chains. Then, one can still identify manifolds with different
number of excitations, as the gap between bands is larger than
energy differences within each band. However, for large $\delta$ and
for long chains the spectrum becomes more involved, and the gaps
between different excitation states diminish. In what follows we
restrict ourselves to situations where the gaps between manifolds
are maintained, $\sim \epsilon$, larger than the spacings within
each band (bandwidth $\sim \kappa$). Practically, we study chains of
$N<10$ units, with large onsite gaps $\epsilon\gg\kappa$ and small
anisotropy parameter $0<\delta<1$.

The spin chain serves as a prototype model for exploring energy
transfer through a linear molecular junction, at low temperatures.
In what follows we study the steady state behavior of this model in
two different limits:

(i) Off-resonance case, with $\epsilon \gg \omega_c$ or
 $T_{\nu}<\epsilon$. Here
$\omega_c$ is the reservoirs cutoff frequency. In this limit
energy transfer takes places between low frequency
modes of the two reservoirs, mediated by a (high frequency)
subsystem. For treating the dynamics in this scenario, we adopt the
recently developed Energy-Transfer Born-Oppenheimer method
\cite{BOheat}. In Sec. III we show that in this regime heat is
transferred via a coherent-superexchange mechanism, with the current
exponentially decreasing with chain size.

(ii) Resonance regime, with  $\epsilon < \omega_c$ and $T_{\nu}>
\epsilon$. Under these conditions baths' modes in resonance with the
subsystem frequencies are populated, responsible for the subsystem
excitation and relaxation processes. This resonance energy transfer
process can be treated within the Born-Markov approximation scheme
\cite{MasterD,Hyblong}. In Sec. IV we show that in this case the isotropic XY
model transfers energy in a ballistic manner.

\section{Off-resonance regime: Energy-Transfer Born-Oppenheimer scheme}

\subsection{Method}
We describe the principles of the Energy-Transfer Born-Oppenheimer
method as developed in Ref. \cite{BOheat}, then apply it onto the
spin-chain model, to obtain the behavior of the current as a
function of size. Generally, the BO approximation \cite{BO} is based
on the recognition of timescale separation. In isolated molecules,
the "traditional" BO approximation relays on the mass separation of
electrons and atomic nuclei. In this context, one assumes that the
electron cloud instantly adjusts to changes in the nuclear
configuration, and that the nuclei propagate on a single potential
energy surface associated with a single electronic quantum state,
obtained by solving the Schrodinger equation with fixed nuclear
geometries.

This principle can be adopted for treating quantum thermal transport
in (potentially strong) interacting systems driven to a steady state
by a temperature bias \cite{BOheat}. The method is applicable in the
{\it off-resonant} regime, where the characteristic frequencies of
the impurity object are high relative to the cutoff frequencies of
the reservoirs $\epsilon \gg \omega_c$ \cite{comment}.
This implies a timescale separation, as the subsystem dynamics is
fast, while the bath motion is slow.  The ETBO approximation follows
two consecutive steps: First, we consider the fast variable and
solve the subsystem eigenproblem while fixing the reservoirs
configuration, to acquire a set of potential energy surfaces which
parametrically depend on the bath coordinates. In the second step we
adopt the adiabatic approximation and assume that the baths'
coordinates, the slow variables, evolve without changes on the
subsystem state. We then solve the energy transfer problem between
the reservoirs on a fixed potential surface.

In what follows we denote by $q$ subsystem coordinates and by
$Q_{\nu}$ the $\nu$ bath coordinates. These are collections of
displacements and momenta operators. The baths operators which are
coupled to the subsystem, $B_{\nu}$, are functions of the $Q_{\nu}$
coordinates. We also collect in $Q$ the coordinates of both
reservoirs. Fixing the bath coordinates, we identify the 'fast'
contribution to Eq. (\ref{eq:H}) as
\bea
H_{f}(q,Q)=H_{S}(q)+ \sum_{\nu}V_{\nu}(q,Q_{\nu}),
\label{eq:Hg}
\eea
and solve the time-independent Schr\"odinger equation
\bea
H_f(q,Q)|g_n(q,Q)\rangle = W_n(Q)|g_n(q,Q)\rangle,
\label{eq:diag}
\eea
to acquire a set of "potential energy surfaces", $W_n(Q)$, and states
$|g_n(q,Q)\rangle$. Note that the potentials $W_n$ mix the left and
right system-bath interaction operators. Moreover, they are not necessarily
linear in $Q_L$ and $Q_R$.
These potentials are the analogs of the electronic potential energy surfaces
obtained in molecular structure calculations, which parametrically depend on
the nuclear coordinates.
Similarly, the reservoir coordinates $Q$ are treated as parameters in Eq. (\ref{eq:diag}).
Assuming that the surfaces are well
separated, we presume the adiabatic ansatz and write the total density
matrix as
\bea
\rho(t)=\left\vert g_n(q,Q)\right\rangle
\rho_{B}^n(Q,t)
\left\langle
g_n(q,Q)\right\vert,
\label{eq:rho}
\eea
where the bath density matrix obeys the Liouville equation
\bea
\rho_B^n(Q,t)= e^{-iH_{BO}^n t}\rho_B(0)  e^{iH_{BO}^n t},
\label{eq:rhoB}
\eea
$\hbar\equiv 1$, with the effective Hamiltonian
\bea
H_{BO}^n=H_{L}(Q_L)+H_{R}(Q_R)+W_n(Q_L,Q_R).
\label{eq:HBO}
\eea
Here $\rho_B(0)=\rho_L \times \rho_R$ is a factorized initial
condition with $\rho_{\nu}=e^{-\frac{H_{\nu}}{T_{\nu}}}/{\rm
Tr_{\nu}}\big[e^{-\frac{H_{\nu}}{T_{\nu}}}\big]$, the
equilibrium-canonical distribution function of the $\nu$ bath. In
what follows, we assume that the baths coordinates evolve on the
{\it ground potential surface}, simply denoted by $W$. The effective
Hamiltonian (\ref{eq:HBO}) including the ground potential surface
$W$ will be similarly denoted by $H_{BO}$.
For brevity, we also omit references to coordinates.
Our plan is to study next the quantum dynamics dictated by the Hamiltonian
(\ref{eq:HBO}), on a particular surface. Such an analysis is analogous to the investigation
of vibrational dynamics on a particular electronic potential surface,
in the traditional application of the BO approximation.



In steady state, the energy current operator, e.g., at the $L$
contact, can be defined as \cite{Wucurr}
\bea
\hat J_{L}=i[H_{L},W],
\label{eq:Jop}
\eea
with the expectation value
\bea J_{L}(t) =\text{Tr}[\hat J_{L}\rho _{B}(t)]
=\text{Tr}[e^{iH_{BO}t}\hat J_{L}e^{-iH_{BO}t}\rho_{B}(0)].
\label{eq:Jav} \eea
The left expression is written in the Schr\"odinger picture; the
second is in the Heisenberg representation. The trace is performed
over the two baths' degrees of freedom. When system-baths couplings,
absorbed into $W$, are weak, the time evolution operator can be
approximated by the first order term
\bea
e^{-iH_{BO}t}=e^{-i(H_{L}+H_{R})t}\left(1-i\int_{0}^{t}W(\tau )d\tau \right),
\label{eq:weak}
\eea
and the current (\ref{eq:Jav}) reduces to 
\bea J_{L}(t)=-i\int_{0}^{t}\text{Tr}\{[\hat J_{L}(\tau
),W]\rho_{L}\rho_R\}d\tau. \label{eq:curr} \eea
Here $W(\tau)$ and $\hat J_{L}(\tau)$ are interaction picture
operators, $O(t)=e^{iH_Bt}O e^{-iH_Bt}$ with $H_B=H_L+H_R$. We are
interested in  steady-state quantities, $J=J_{L}(t\rightarrow \infty
)$, if the limit exists. Expression (\ref{eq:curr}) can be further
customized, recalling the bipartite interaction form of $V_{\nu}$ in
the original Hamiltonian (\ref{eq:H}). Then, one can formally expand
$W$ in terms of the bath operators which are coupled to the
subsystem, $B_{\nu}$,
\bea W&=&\sum_{a,b}  A_{a,b}B_L^{a}\otimes B_R^{b} \label{eq:W}
\\
&=&\sum_{a,b}\sum_{k,m} \sum_{p,s}  A_{a,b} (B_L^{a})_{km}
(B_R^{b})_{ps}\left\vert kp\right\rangle \left\langle m
s\right\vert. \nonumber \eea
The operators $B_L$ and $B_R$ depend on the bath coordinates,
collected into $Q_L$ and $Q_R$, respectively. The actual form is not
important at this (formal) stage. It is specified only once
particular models are constructed, see e.g., Eq. (\ref{eq:Harm}).
The coefficients $A_{a,b}$ absorb the subsystem parameters, the
energies $\epsilon$, $\kappa$  and $\delta$ in the chain model and
the system-bath interaction strength $\lambda_{\nu}$. The powers $a$
and $b$ are positive integers. $|k\rangle$ and $|m\rangle$ represent
the many body states of the left reservoir with energies $E_k$ and
$E_m$, (i.e. $H_{L}=\sum E_{k}\left\vert k\right\rangle \left\langle
k\right\vert$). Similarly, $|p\rangle$ and $|s\rangle$ are the many
body states of the right reservoir with energies $E_p$ and $E_s$.
Assuming a weak system-bath coupling strength, we truncate $W$ and consider
only the lowest order term in $B_LB_R$,
\bea W\sim A_{0,0}
+A_{1,1}B_LB_R + O(B_L^2) +O(B_R^2).
\label{eq:WW} \eea
A more general derivation in presented in Ref. \cite{BOheat}. It can
be shown that only terms containing products of $B_L$ and $B_R$ add
to the current, thus only the second term in Eq. (\ref{eq:WW})
actually matters for the energy current calculations. We also note
that $A_{1,1}$ is proportional to the product $\lambda_L\lambda_R$,
see Eq. (\ref{eq:HV}). We therefore define the function $\mathcal
T(\epsilon,\kappa)$ through the relation
\bea A_{1,1}\equiv\lambda_L \lambda_R \mathcal T (\epsilon,\kappa),
\label{eq:A11}
\eea
where we explicitly indicate its dependence on the subsystem
parameters. Back to (\ref{eq:curr}), employing Eqs. (\ref{eq:Jop})
and (\ref{eq:WW}), we obtain
\bea J&=& \frac{\mathcal T(\epsilon,\kappa)^2}{Z_LZ_R} \int_0 ^{\infty }dt
 \Big[ \sum_{k,m}\lambda_L^2 E_{km}e^{iE_{km}t}(B_L)_{km}
(B_L)_{mk}e^{-\beta_LE_k}
\nonumber\\
&&\times \sum_{p,s} \lambda_R^2
(B_R)_{p,s}(B_R)_{s,p}e^{iE_{ps}t}e^{-\beta_RE_p} +c.c. \Big],
\eea
where, e.g.,  $Z_{L}=\sum_k e^{-\beta_{L}E_{k}}$
is the $L$ bath partition function; $\beta_{\nu}=1/T_{\nu}$, $k_B\equiv1$, and
$E_{km}=E_{k}-E_{m}$.
Time integration can be readily performed, leading to the steady state heat current
\bea J=\frac{\mathcal T(\epsilon,\kappa)^2}{2\pi}  \int_0^{\infty} \omega
d\omega \big[ k_{L+}(\omega) k_{R-}(\omega)
 - k_{L-}(\omega) k_{R+}(\omega)\big].
 \label{eq:currF}
 \eea
The excitation ($+$) and relaxation ($-$) rate constants are given by
\bea
k_{L\pm }(\omega)=2\pi\sum_{k,m} \lambda_L^2
[(B_L)_{km}(B_L)_{mk}]^{\pm }
\delta(E_k-E_m\mp\omega)\frac{e^{-\beta_LE_k}}{Z_L}. \eea
We have introduced here the short notation $[ (B_L)_{km}
(B_{L})_{mk}]^{+}$, to denote matrix elements when $E_{k}>E_{m}$.
Similarly, $[ (B_L)_{km} (B_{L})_{mk}]^{-}$ describes the
$E_{k}<E_{m}$ case. Analogous expressions hold for the $R$ rates. We
can also rewrite the rate constants as Fourier transforms of bath correlation
functions
\bea k_{L\pm }(\omega) = \lambda_L^2\int_{-\infty}^{\infty}e^{\mp
i\omega t}{\rm Tr}\left[\rho_L B_L(t)B_L(0)\right] dt,
\label{eq:rate} \eea
satisfying detailed balance,
$k_{L+}(\omega)=k_{L-}(\omega)e^{-\beta_L \omega}$. Using
this relation, we organize Eq. (\ref{eq:currF}) as
\bea J=\frac{\mathcal T (\epsilon,\kappa)^2}{2\pi} \int_0^{\infty}
\omega d\omega k_{L-}(\omega) k_{R-}(\omega) (
e^{-\beta_L\omega}-e^{-\beta_R\omega}).
\label{eq:currFF}
\eea
This result is given in the form of a "generalized Landauer
formula": The net heat current is given as the difference between
left-moving and right-moving excitations, nevertheless, unlike the
original Landauer formula \cite{Land}, this expression can
incorporate anharmonic effects within the
chain model, eventually absorbed into $\mathcal T$,
and nonlinear system-bath interactions, taken in by the rates
$k_{\nu\pm}(\omega)$. We emphasize the broad status of Eq.
(\ref{eq:currFF}): It does not assume a particular structure for the
subsystem, or a specific system-bath interaction form, $B_{\nu}$,
both contained inside $W$. It is valid as long as (i) there exists a
timescale separation between the subsystem motion (fast) and the
reservoirs dynamics (slow), and (ii) system-bath interactions, given
in a bipartite form, are weak, see Eqs. (\ref{eq:weak}) and
(\ref{eq:WW}).

Eq. (\ref{eq:currFF}) readily reveals the dependence of the current
on the subsystem parameters, thus it is immensely useful for
exploring transport behavior. It includes a product of two terms:
The prefactor depends on the subsystem parameters, the integral over
frequencies encompasses the bath operators within the Fermi golden
rule rates (possibly nonlinear in the bath coordinates). 
The effect of the reservoirs' temperatures is enclosed there. Since
the prefactor $\mathcal T(\epsilon,\kappa)$ is the only term
corroborating chain parameters, by obtaining the ground state
surface $W$ [Eqs. (\ref{eq:WW})-(\ref{eq:A11})], the scaling of the
current with size and energy can be gained, without solving a
dynamical problem.

We can also regard Eq. (\ref{eq:currFF}) as a generalization of
the nonadiabatic transition rate, $k_{da}=2\pi|V_{da}|^2 FCWD$,
describing electron or energy transfer processes within a donor-bride-acceptor
complex, to current carrying steady state situations. Here, the
Franck-Condon factor $FCWD$ accounting for the conservation of
energy, depends on the temperature of the environment \cite{Scholes}.
In Eq. (\ref{eq:currFF}) this term is portrayed by the frequency
integral, considering a transport process originating from a
particular state within the $L$ bath. The second part, $V_{da}$,
combines the electronic coupling between the donor and acceptor
states. In the present work this factor is accounted for by the
function $\mathcal T(\epsilon, \kappa)$.

As an example of the utility of the ETBO method to describe off-resonance conduction processes,
consider a harmonic impurity of frequency $\Omega$, linearly coupled to two
harmonic reservoirs,
\bea H_S&=& \Omega b^{\dagger}b,
\nonumber\\
H_{\nu}&=&\sum_j \omega_j b_{\nu,j}^{\dagger}b_{\nu,j},\,\,\,\,\,\,\, V_{\nu}= (b^{\dagger}+b)\lambda_{\nu} B_{\nu}
\label{eq:Harm}
\eea
with $B_{\nu}=\sum_j (b_{\nu,j}^{\dagger}+b_{\nu,j})$. Here
$b_{\nu,j}^{\dagger}$ ($b_{\nu,j}$) are the creation (annihilation)
operators of the mode $j$ in the $\nu$ bath, $b^{\dagger}$ and $b$
are the respective subsystem operators. Since the model is fully
harmonic, in principle the energy current can be exactly obtained.
However, this calculation requires some effort, and the scaling of
the current with size is not easy to reveal \cite{SegalHeat,Dhar}.
Focusing on the off-resonance limit, the ETBO method can readily
provide the behavior of the current at weak couplings.
We diagonalize $H_f=H_S+V$ and resolve the
ground potential surface $W= -\frac{2}{\Omega}
(\lambda_LB_L+\lambda_RB_R)^2$, thus extract $\mathcal T \propto
1/\Omega$. Relaying on the bilinear interaction form, the transition
rates (\ref{eq:rate}) can be  obtained, $k_{\nu+}(\omega)=
\Gamma_{\nu}(\omega) n_{\nu}(\omega)$, where
$n_{\nu}(\omega)=[e^{\beta_{\nu}\omega}-1]^{-1}$ is the
Bose-Einstein distribution function and the coefficient
$\Gamma_{\nu}(\omega)$ incorporates the system-bath interaction
strength and the bath's density of states, assumed to be weak
$\Gamma_{\nu}(\omega)<\Omega$. Combining these elements in the
expression for the heat current (\ref{eq:currFF}), we conclude that
\bea J\propto \frac{1}{\Omega^2}\int_0^{\infty} \omega d\omega
\Gamma_L(\omega)\Gamma_R(\omega) [n_L(\omega)-n_R(\omega)]. \eea
To be consistent with the off-resonance assumption, one should
evaluate this expression at low temperatures $T_{\nu}<\Omega$, or
impose a cutoff for the reservoirs frequencies, $\Omega\gg\omega_c$.
This result exposes the scaling of the current with the
subsystem energy and the baths temperatures. It can be shown that
similar scaling holds for the spin-boson model in the off-resonance
limit \cite{BOheat,Teemo}. This correspondence is physically correct
since at low temperature an harmonic impurity behaves similarly to a
spin impurity, as transport takes place through the lowest
excitations of the subsystem.

\subsection{Analytic Results}
We apply the ETBO formalism on the spin-chain Hamiltonian
(\ref{eq:H})-(\ref{eq:HV}), to obtain the energy current
characteristics. Our objectives are (i) to resolve the behavior of
the current as a function of chain size, and (ii) to obtain its
dependence on the subsystem energetics, $\epsilon$ and $\kappa$.
With this at hand, we can identify the dominant transport mechanism.
For simplicity, we exemplify our analysis using the isotropic XY
chain model. However, the results are applicable for other models
including the anisotropic Heisenberg model as well, for $\delta<1$,
see discussion below Eq. (\ref{eq:HspinHeis}). We comment on this
model below Eq. (\ref{eq:Texp}).

We recall that the basic ingredient of the ETBO formalism is the ground
potential surface $W$, or its expansion, (\ref{eq:WW}). Then,
identifying the coefficient $\mathcal T(\epsilon,\kappa)$, the
energy and size dependent of the current can be captured using Eq.
(\ref{eq:currFF}).
We review the elements of our model introducing a more compact notation for
the chain subsystem,
\bea
H_S=\epsilon \hat M+\kappa \hat h,
\label{eq:HZ}
\eea
with $\hat M=\sum_{j=1}^N \sigma _{j}^{+}\sigma _{j}^{-}$
 and $\hat h$ as the hopping Hamiltonian, including nearest-neighbor interactions
 \cite{comment2}. For example,
 $\hat h=\frac{1}{2}\sum \left(
\sigma_j^x\sigma_{j+1}^x+\sigma_j^y\sigma_{j+1}^y \right)$. The
chain is connected by $V_{\nu}$ to the thermal bath $H_{\nu}$. The
total Hamiltonian is given by
\bea H&=&H_f+H_L+H_R,
\nonumber\\
H_f&=&H_S+V, \,\,\,\,\,\,\,\, V=V_L+V_R, \label{eq:HR} \eea
with $V_L=S_1B_L$ and $V_R=S_NB_R$; $S_{1,N}\propto \sigma_{1,N}^x$ contains subsystem
operators. The energy surface $W$, the lowest
eigenenergy of $H_f$, is  accomplished through the eigenvalue equation
\bea H_f|g_0\rangle =W |g_0\rangle. \eea
Since an exact diagonalization is limited to simple models
\cite{BOheat}, in this work we construct $W$ using time independent perturbation theory.
As the unperturbed basis we utilize the subsystem eigenstates $|n\rangle$, satisfying
\bea H_S|n\rangle =E_n |n \rangle. \eea
The system-bath interaction operator $V$  plays the role of a  perturbation.
These  $|n\rangle$ states include different number of excitations, demonstrated in Fig.
\ref{Fig4}. For example, for a two-qubit chain,
\bea H_{S}=\epsilon (\sigma _{1}^{+}\sigma _{1}^{-}+\sigma
_{2}^{+}\sigma _{2}^{-})+\frac{\kappa }{2}(\sigma _{1}^{x}\sigma
_{2}^{x}+\sigma _{1}^{y}\sigma _{2}^{y}), \eea
we obtain the eigenfunctions and respective energies
\bea \left\vert 0\right\rangle &=&\left\vert \downarrow \downarrow
\right\rangle , \,\,\,E_{0}=0
\nonumber\\
\left\vert 1\right\rangle &=&\frac{1}{\sqrt{2}}(\left\vert
\downarrow \uparrow \right\rangle -\left\vert \uparrow \downarrow
\right\rangle ),\,\,\,E_{1}=\epsilon -\kappa
\nonumber\\
\left\vert 2\right\rangle &=&\frac{1}{\sqrt{2}}(\left\vert
\downarrow \uparrow \right\rangle +\left\vert \uparrow \downarrow
\right\rangle ),\,\,\,E_{2}=\epsilon +\kappa
\nonumber\\
\left\vert 3\right\rangle &=&\left\vert \uparrow \uparrow
\right\rangle ,\,\,\,E_{3}=2\epsilon. \label{eq:2qbit} \eea
The ground state is fully polarized, $|\downarrow\downarrow
\rangle$, with the two spins in their ground state. The first two
excited states include a single excitation (a superposition,
residing on the first and second sites). The high energy state includes
two excitations.
Back to the $N$-site chain, the ground state energy of $H_f=H_S+V$
can be written by using time independent perturbation theory to the
second order correction,
\bea W\sim E_{0}+ \langle 0| V|0 \rangle+ \sum_{n\neq
0}\frac{\left\vert \left\langle 0\right\vert V\left\vert
 n\right\rangle \right\vert ^{2}}{E_{0}-E_{n}}.
 \label{eq:Wp}
\eea
The corresponding eigenfunction is
\bea
|g_0\rangle \sim \left\vert 0\right\rangle +\sum_{n\neq 0}\frac{\left\langle 0\right\vert
V\left\vert n\right\rangle }{E_{0}-E_{n}}\left\vert n\right\rangle.
\label{eq:g0}
\eea
Consider now a family of spin Hamiltonians where the ground state is
fully polarized as in (\ref{eq:2qbit}), $\left\vert 0\right\rangle
=\left\vert \downarrow,\downarrow,...,\downarrow\right\rangle $. The
structure of the system-bath interaction operator, $\sigma^x_{1,N}$,
allows to connect this ground state  only to
single excitation states, the subgroup $|n_1\rangle \in |n\rangle$, 
written as
\bea \left\vert n_1\right\rangle =\sum_{j=1}^{N}C_{j}^{n}\left\vert
j\right\rangle. \label{eq:n}
\eea
These are linear combinations of single excitation states,
$\left\vert j\right\rangle =\left\vert
\downarrow,\downarrow,.,\uparrow_{j},..,\downarrow\right\rangle$.
The eigenenergies of these states are $E_{n_1}=\epsilon + \kappa
\alpha_{n_1}$ with $\alpha_{n_1}$ as numerical coefficients. For example, $\alpha_{n_1}=\pm1$ for the 2-qubit
chain of Eq. (\ref{eq:2qbit}).
Identifying the relevant states $|n_1\rangle$, we now plug them
and their corresponding energies into Eq. (\ref{eq:Wp}).
The constant shift $E_0$ was set here to zero,
the second term vanishes.
Therefore, the ground potential surface is given by
%
\bea &&W\sim \sum_{n_1}\frac{\left\vert \left\langle 0\right\vert
V\left\vert
 n_1\right\rangle \right\vert ^{2}}{E_{0}-E_{n_1}}
 \nonumber\\
&& =
 -\frac{B_LB_R}{\epsilon }
 \sum_{n_1}\frac{\left\langle
0\right\vert \sigma _{1}^{x}\left\vert n_1\right\rangle \left\langle
n_1\right\vert \sigma _{N}^{x}\left\vert 0\right\rangle
+\left\langle 0\right\vert \sigma _{N}^{x}\left\vert
n_1\right\rangle \left\langle
n_1\right\vert \sigma _{1}^{x}\left\vert 0\right\rangle }{1+\frac{\kappa }{%
\varepsilon }\alpha_{n_1}} +O(B_L^2)+O(B_R^2).
\label{eq:W1}
\eea
Terms which involve only $B_L$ or $B_R$ operators do not contribute
to the current and are therefore ignored. Focusing on the sum,
denoted by $S$, we simplify it recalling that $\kappa<\epsilon$. We
expand the denominator using the geometric sum formula,
$\sum_{q=0}^{\infty} x^q=\frac{1}{1-x}$,
 \bea S&=&
\sum_{n_1,q}\left(-\frac{\kappa }{\epsilon }\alpha
_{n_1}\right)^{q}\left[\left\langle 1|n_1\right\rangle \left\langle
n_1|N\right\rangle +\left\langle N|n_1\right\rangle \left\langle
n_1|1\right\rangle\right]
\nonumber\\
&=&\sum_{n_1,q}\left(-\frac{\kappa }{\epsilon }
\right)^{q}\left[\left\langle 1\right\vert \hat h^{q}\left\vert
n_1\right\rangle \left\langle n_1|N\right\rangle +\left\langle
N\right\vert \hat h^{q}\left\vert n_1\right\rangle \left\langle
n_1|1\right\rangle \right]
\nonumber\\
&=&\sum_{q}\left(-\frac{\kappa }{\epsilon }
\right)^{q}\left[\left\langle 1\right\vert \hat h^{q}\left\vert
N\right\rangle +\left\langle N\right\vert \hat h^{q}\left\vert
1\right\rangle \right]. \label{eq:D1} \eea
%
Here, the states  $\left\vert 1\right\rangle$ and $\left\vert
N\right\rangle$ refer to a $j$-type state as defined below Eq.
(\ref{eq:n}), containing a single excitation in the leftmost (1) site or in
the rightmost ($N$) site.
The second line was derived using the eigenequation for the hopping operator, $\hat h |n_{1}\rangle=
\alpha_{n_1}|n_{1}\rangle$.
The last line was obtained by using the
fact that $\langle n_j|N\rangle=0$ and $\langle n_j|1\rangle=0$ for $j>2$,
where $|n_j\rangle$ denotes states
with $j$ excitations residing on the chain. The
completeness relation is also invoked, $I=\sum|n\rangle\langle n|$.

We can further simplify Eq. (\ref{eq:D1}). We note that $\hat h$ is
the inter-site hopping operator, and use the fact that it includes
nearest-neighbor interactions only. This leads to $\left\langle
1\right\vert \hat h^{q}\left\vert N\right\rangle =0$ if $q<(N-1)$.
Therefore, the leading term of the $q$ expansion must include $N-1$
operators for transferring an excitation from the first  unit of the chain
to the last one,
\bea
W\sim
-\frac{1}{\epsilon }\left(-\frac{\kappa }{\epsilon } \right)^{N-1}
 B_{L}B_{R}
\left[ \left\langle 1\right\vert \hat h^{N-1}\left\vert
N\right\rangle +\left\langle N\right\vert \hat h^{N-1}\left\vert
1\right\rangle  \right].
 \eea
The square brackets yield a numerical factor. We conclude
that the ground state potential follows a simple form
\bea W\sim \mathcal T (\epsilon,\kappa) B_LB_R, \label{eq:Wt} \eea
with
\bea \mathcal T (\epsilon,\kappa)=-\frac{1}{\epsilon}\left
(-\frac{\kappa}{\epsilon} \right)^{N-1}. \label{eq:At} \eea
We now go back to the energy current (\ref{eq:currFF}), denoting by
$F(T_{L},T_R)$ the contribution that depends on the reservoirs'
temperatures,
\bea J=
\frac{1}{\epsilon^2}\left(\frac{\kappa}{\epsilon}\right)^{2N-2}
F(T_L,T_R). \label{eq:tunn1} \eea
One can also express the prefactor by a decaying exponent,
\bea \mathcal T(\epsilon,\kappa)\propto e^{-\alpha N}, \,\,\,\,
\alpha=-2\ln(\kappa/\epsilon).
\label{eq:Texp}
\eea
Eq. (\ref{eq:tunn1})
describes an exponential decay of the energy current with distance, a
coherent-superexchange dynamics \cite{superEx,EET}. The physical
picture exposed is that low frequency (reservoirs) modes are being
coherently exchanged, without the actual excitation of the
(off-resonance) modes of the chain. The intermediating structure-chain
therefore serves as a mediating medium, allowing for through-bond
couplings. This expression is the analog of the McConnell
super-exchange result, describing deep electron tunneling in tight
binding models \cite{superEx}.


\begin{figure}
{\hbox{\epsfxsize=80mm\epsffile{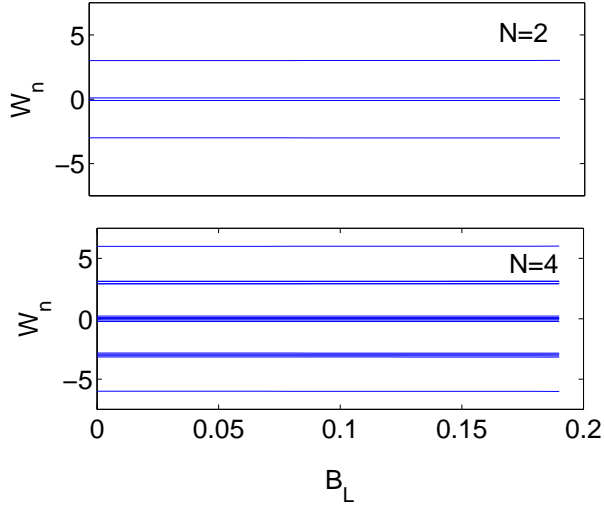}}} \caption{Potential energy
surfaces for $N=2$ (top) and $N=4$ (bottom), $\epsilon=3$,
$\kappa=0.2$, $\delta=0$, $B_R=0.1$.} \label{Fig0}
\end{figure}

\begin{figure}
{\hbox{\epsfxsize=80mm\epsffile{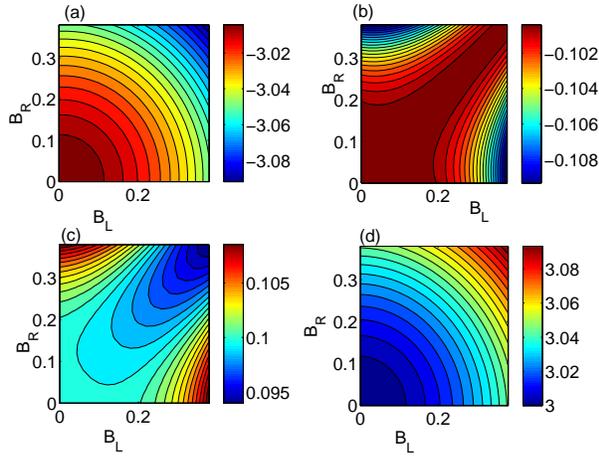}}} \caption{
Contour plot of the four
potential energy surfaces for $N=2$. Panels
(a) to (d) present the potentials from the ground state upward,
$\epsilon=3$, $\kappa=0.2$, and $\delta=0$.}
\label{FigW3}
\end{figure}

The derivation above is applicable for models more general than the
Hamiltonian (\ref{eq:H})-(\ref{eq:HV}). For example, we may modify
the spin chain and add an interaction term $\delta \sigma_j^z
\sigma_{j+1}^z$; The exponential result still holds as we show next.
Overall, the analysis relays on the following ingredients: (i) the
ground state of $H_S$ is a fully polarized state, i.e., there are no
excitations on the bridge in the absence of the interaction with the
reservoirs. (ii) system-bath interactions involve the generation of
a single excitation on the chain boundary sites, and (iii) the chain
units are weakly connected through nearest-neighbor couplings, small
relative to the onsite gap, $\kappa\ll\epsilon$.
Under these conditions, combined with the perquisites for the
validity of the ETBO approach (off-resonance condition and weak
system-bath couplings), transport dynamics  reflects the superexchange mechanism, Eq.
(\ref{eq:Texp}).

The above analysis holds, under some conditions, for describing the dynamics of
the anisotropic Heisenberg chain in the off-resonance regime.
For a 2-qubit model
\bea
H_{S}=\epsilon \left(\sigma _{1}^{+}\sigma _{1}^{-}+\sigma
_{2}^{+}\sigma _{2}^{-}\right)+\frac{\kappa }{2}\left(\sigma _{1}^{x}\sigma
_{2}^{x}+\sigma _{1}^{y}\sigma _{2}^{y}+\delta \sigma _{1}^{z}\sigma
_{2}^{z}\right). \eea
%
We repeat the derivation, Eqs. (\ref{eq:W1})-(\ref{eq:D1}), and resolve the ground potential
\bea
&&W= \frac{(B_{L}-B_{R})^{2}}{2(-\epsilon +\kappa +\kappa \delta )%
}+\frac{(B_{L}+B_{R})^{2}}{2(-\epsilon -\kappa +\kappa \delta )}
\nonumber\\
&&=
-\frac{1}{(\epsilon -\kappa \delta )^{2}-\kappa ^{2}}
\left[ \left(\epsilon -\kappa \delta\right) B_{L}^{2}-2\kappa B_{L}B_{R}+\left(\epsilon -\kappa
\delta \right)B_{R}^{2}\right ]
\nonumber\\
&\sim&  \left[2\frac{\kappa }{\epsilon ^{2}}+4\delta \frac{ \kappa
^{2}}{ \epsilon ^{3}}+\allowbreak 2\frac{\kappa ^{3}}{\epsilon
^{4}}\left( 1+3\delta ^{2}\right)
\right]B_{L}B_{R}
+ O(B_L^2)+O(B_R^2).
\eea
The last line was derived under the weak exchange assumption,
$\epsilon\gg\kappa$. This result agrees with the behavior predicted
in Eqs. (\ref{eq:Wt})-(\ref{eq:At}) when $\delta=0$. We also note
that the exchange anisotropy parameter $\delta$ affects $W$ to
higher order in $\kappa/\epsilon$. Thus, to the lowest order in
$\kappa/\epsilon$ the energy current satisfies $J(N=2)\propto
\frac{\kappa^2}{\epsilon^4}F(T_L,T_R)$, irrespective of the details
of the spin model. This behavior prevails for longer chains as well:
The onset of $\delta$ provides only higher order corrections to the
off-resonant energy current (\ref{eq:tunn1}) when $\delta$ is small.
More precisely, the results hold as long as the spectrum maintains
its distinct band structure, see Fig. \ref{Fig4a}, the (single)
excitation energies could be still approximated by
$E_{n_1}\sim\epsilon+\kappa\alpha_{n_1}$, and the ground state is
fully polarized.


We now comment on the usefulness of the ETBO method to describe
transient effects. While the approach has been formulated for
treating non-equilibrium steady state situations \cite{BOheat}, one
could also rewrite it to describe  the {\it transient} dynamics of a
donor excitation, transferred to an acceptor sidegroup through a
bridging backbone. In the context of electron transfer, twp distinct
quantities, the electrical conduction and the electron transfer
rate, were shown to be linearly related \cite{Nitzan}. Similar
correspondence should arise in the context of excitation energy
transfer \cite{EET}, comparing the steady state energy current at
very small temperature bias and the excitation transfer rate. We
thus argue that the thermal conductance, obtained as $\lim_{\Delta T
=0}J/\Delta T$, is proportional to the excitation energy transfer
rate detected in donor-bridge-acceptor complexes \cite{Nitzan}.


\begin{figure}
{\hbox{\epsfxsize=80mm\epsffile{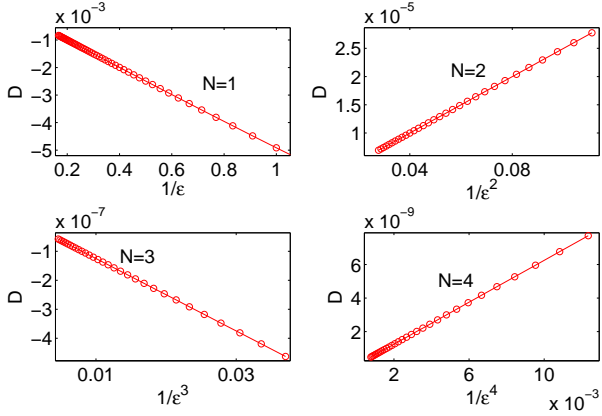}}} \caption{The function
$\mathcal D=B_LB_Rf_S(\epsilon,\kappa)$ for $N=1,2,3,4$. The bath
coordinates are fixed at $B_L=B_R=0.05$. The exchange parameters are
$\kappa=0.1$, $\delta=0$.}
\label{Fig1}
\end{figure}

\begin{figure}
{\hbox{\epsfxsize=80mm\epsffile{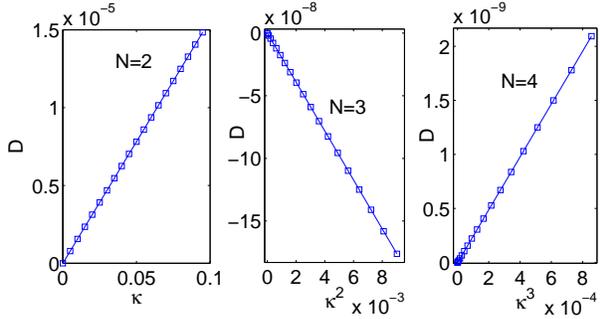}}} \caption{The function
$\mathcal D=B_LB_Rf_S(\epsilon,\kappa)$ for $N=2,3,4$. The bath
coordinates are fixed at $B_L=B_R=0.05$. The exchange parameters
are $\epsilon=4$, $\delta=0$.}
\label{Fig2}
\end{figure}


\subsection{Numerical Simulations}
We support our analysis by an exact numerical
diagonalization of $H_f$, to obtain the set of potential surfaces $W_n$.
We recall that the potential surfaces $W_n(Q)$, with $Q$ enclosing the bath coordinates coupled to the system,
are the analogs of the electronic surfaces in the context of molecular structure,
generated by varying the (slow) nuclear coordinates.
Here, in a similar fashion we treat the bath coordinates $Q$
as parameters, overall treating $B_L(Q)$ and $B_R(Q)$ as parameters.
The model consists the isotropic XY spin chain coupled at the boundaries to two thermal
reservoirs, Eqs. (\ref{eq:H})-(\ref{eq:HV}).
First, in support of the adiabatic approximation we show in Fig. \ref{Fig0} that the
ground potential energy surface $W$ is separated by a
substantial gap from other states. The $x$-axis is the bath
coordinate $B$. Practically, the data was generated by fixing $B_R$
and parametrically modifying the coordinate $B_L$. The ground potential surface $W$ lies
around $-N\epsilon/2$. It is separated by $\sim \epsilon$ from the
higher excitation states. This observation consistently supports the
BO scheme. 
While in Fig. \ref{Fig0} the potentials seem flat due to the scale used,
in Fig. \ref{FigW3} we explicitly present a contour plot of the four potential
surfaces $W_n$ for $N=2$, to demonstrate their variance with the thermal bath coordinate $B_{\nu}$.
We now select $W$, the ground potential, and postulate that it
fulfills Eq. (\ref{eq:WW}), or more generally,
\bea W&=&f_0(\epsilon) + f_L(B_L^2,\epsilon, \kappa) +
f_{R}(B_R^2,\epsilon, \kappa)
\nonumber\\
&+& B_LB_R f_S(\epsilon,\kappa) +O(B_L^2B_R^2),
\label{eq:Eg}
\eea
with the unknown functions $f$. For resolving the part in $W$
responsible for transport, we define an auxiliary function
\bea W'=W-W(\kappa=0). \eea
The difference $\mathcal D\equiv W'-W'(B_L=0)-W'(B_R=0)$ should
isolate the (fourth) term in Eq. (\ref{eq:Eg}), the term which
contributes to the energy current in the lowest order of the
system-bath interaction strength. Figs. \ref{Fig1}-\ref{Fig2}
display this function for chains with $N=$1,2,3, and 4 units
We conclude that the exponential law, Eq. (\ref{eq:At}), is indeed satisfied.
Fig. \ref{Fig1} verifies the exponential dependence on
$\epsilon$, whereas Fig. \ref{Fig2} proves the same for the intersite
coupling $\kappa$.

\section{Resonance transport: Master Equation formalism}

\subsection{Method}

Our objective here is to simulate {\it resonant} energy transfer across isotropic XY
spin chains, assuming that the bath populated modes overlap with the
subsystem gaps. The dynamics is investigated using the Born-Markov
approximation, a second order perturbation theory scheme which
invokes the Markov approximation \cite{Breuer}.
Furthermore, using the secular approximation (SA), the diagonal and nondiagonal
elements of the density matrix are separated.
This scheme results in a markovian quantum master equation.
The method is detailed in Ref. \cite{Hyblong} where it was applied onto an
impurity single-site model. Here, we generalize this treatment for
studying the energy current behavior in an extended system.
Comments about the validity of this approach, to describe  energy transfer processes
in spin chains, are included below Eq. (\ref{eq:rateM}).

We begin by diagonalizing the subsystem Hamiltonian
\bea H_S=L \tilde H_S L^{\dagger}. \eea
$L$ is a unitary matrix which diagonalizes $H_S$. As before, we denote the
resulting eigenspectrum by $|n\rangle$ with $E_n$. The subsystem
operators coupled to the bath are transformed to
the new basis,
\bea \tilde S_1= L^{\dagger} \sigma_1^x L, \,\,\,\, \tilde S_N= L^{\dagger}
\sigma_N^x L. \eea
The operators $\tilde S$ can be formally presented by their
matrix elements as
\bea \tilde S_1&=&\sum_{nm}\tilde S_{1,mn}|m\rangle \langle n|
\nonumber\\
\tilde S_N&=&\sum_{nm}\tilde S_{N,mn}|m\rangle \langle n|. \eea
Since the system is uniform it can be shown that $|S_{1,mn}|^2=|S_{N,mn}|^2\equiv |S_{mn}|^2$, where
the site index is neglected.
The total Hamiltonian in the subsystem basis is  given by
\bea H=\tilde H_S+\lambda_L\tilde S_1B_L+\lambda_R\tilde S_NB_R +H_L
+H_R.
\eea
Under the Born-Markov scheme \cite{Breuer}, accompanied by the SA, the probability to
occupy the $n$ subsystem state can be described by a first order differential
equation
\bea \dot P_n=\sum_{\nu,m} |S_{mn}|^2 P_m(t) k_{m\rightarrow
n}^{\nu} - P_n(t) \sum_{\nu,m} |S_{mn}|^2k_{n\rightarrow
m}^{\nu}. \label{eq:mas1} \eea
The transition rate constants satisfy \cite{Breuer}
\bea k_{m\rightarrow
n}^{\nu}=\lambda_{\nu}^2\int_{-\infty}^{\infty}dt
e^{-iE_{nm}t}{\rm Tr}_B[B_{\nu}(t)B_{\nu}(0)], \eea
where $E_{nm}=E_n-E_m$ and the operators are written in the
interaction representation,
$B_{\nu}(t)=e^{iH_{\nu}t}B_{\nu}e^{-iH_{\nu}t}$. The trace is
performed over the $L$ and $R$ bath states. In steady state, the set
(\ref{eq:mas1}) reduces into a linear set of equations. Complemented
by the conservation of the total probability, $\sum P_n=1$, we can
numerically obtain the steady state occupation probabilities at each
state. The energy current, at the level of the Born-Markov
approximation, is given by \cite{Hyblong}
\bea J=\frac{1}{2} \sum_{n,m}E_{mn} |S_{mn}|^2 P_n (k_{n\rightarrow
m}^{L}- k_{n\rightarrow m}^{R}). \eea
It can be readily calculated once
the steady state population and rate constants are known.
At this stage one should choose a
particular form for the bath operators coupled to the subsystem. For example,
selecting the displacement operators \cite{Hyblong}, the
rate constants reduce to ($m>n)$
 \bea k_{n\rightarrow m}^{\nu} &=&\Gamma(E_{mn}) n_{\nu}(E_{mn})
\nonumber\\
k_{m\rightarrow n}^{\nu} &=&\Gamma(E_{mn})
[n_{\nu}(E_{mn})+1],
\label{eq:rateM}
 \eea
with $\Gamma(\omega)=2\pi\lambda_{\nu}^2\sum_j
\delta(\omega-\omega_j)$. In practice, we take $\Gamma$ as a
constant, independent of frequency, identical at the two contacts.
The function $n_{\nu}(\omega)=[e^{\omega/T_{\nu}}-1]^{-1}$ is the
Bose-Einstein occupation factor.

The authors of Ref. \cite{Michel07} questioned the validity of a
related approach, the Redfield equation, derived in the chain-local
basis, for describing the dynamics of several spin chain systems. In
particular, under the secular approximation, zero energy current was
obtained in  nonequilibrium situations \cite{Michel07}.
Inconsistencies of the Redfield equation, unable to properly reproduce
equilibrium and nonequilibrium dynamics, were noted in the past in
the context of electron transfer processes, see e.g., Ref.
\cite{SegalET}. There, it was argued that working in the subsystem
eigenbasis should lead to a proper equilibration process and to the
correct nonequilibrium dynamics. In view of the zero-current at
finite bias anomaly \cite{Michel07}, we work here in the chain
eigenbasis, indeed naturally eliminating such a nonphysical behavior.

The authors of Ref. \cite{Michel07} further traced the nonphysical
dynamics within the Redfield approach to the inconsistency of the
secular approximation, when applied onto the chain model. It is
argued, that this approximation, resulting in the separation between
the diagonal and nondiagonal terms of the reduced density matrix,
relays on the assumption that differences between the subsystem  energies
are {\it large} compared to the subsystem relaxation rate constants.
However, in the chain model differences between energy states within
each band diminish for long chains, thus one should carefully review
the SA, as we do next.

The eigenspectrum of the isotropic XY spin chain was presented in
Fig. \ref{Fig4}. We recall that for $\epsilon\gg\kappa$ the
subsystem's energies are grouped into manifolds, each containing a
particular number of excitations. It should be noted that the gaps
between bands are preserved, order of $\epsilon$, even for long
chains. We argue next that even though within each manifold the
states become quite dense, one could obtain the correct dynamics of
the isotropic XY model using standard quantum master equation
approaches, stating the SA, as long as the gaps between different
bands are maintained. The reasoning is that once we work in the
chain-diagonal basis, the equation of motion for the density matrix
(before the SA) connects only states which differ by exactly one
excitation through bath excitation and relaxation processes.
Rephrased, states within the same manifold are not directly linked,
only through higher-order bath correlation functions. Thus, within
the Born approximation, energy differences that come into play
within the density matrix equations are always order of the gap
$\epsilon$. Since we pick small relaxation rate constants
$\Gamma<\epsilon$, we conclude that the SA is consistent in the
present setup. This argument does not hold for the Heisenberg model,
as the excitation gaps rapidly disappear with increasing size, see
Fig. \ref{Fig4a}.

\begin{figure}
{\hbox{\epsfxsize=80mm\epsffile{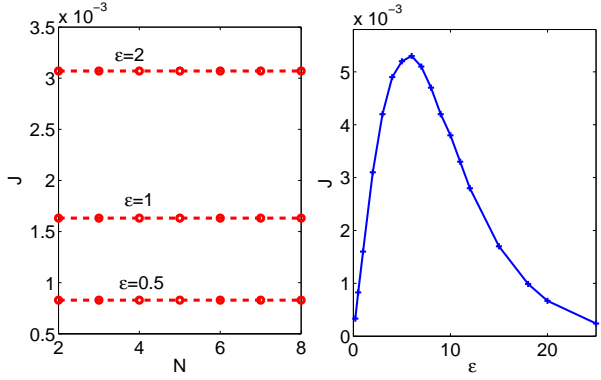}}} \caption{Left panel:
Energy current as a function of size for the isotropic XY chain in the
resonant limit, $\epsilon=$ 0.5, 1, and 2, bottom to top. Right
panel: Energy current as a function of spin gap for $N=6$. Other
parameters are $\kappa=0.1$, $\Gamma=0.01$, $T_L=4$ and $T_R=2$. }
\label{Fig3}
\end{figure}

\begin{figure}
{\hbox{\epsfxsize=80mm\epsffile{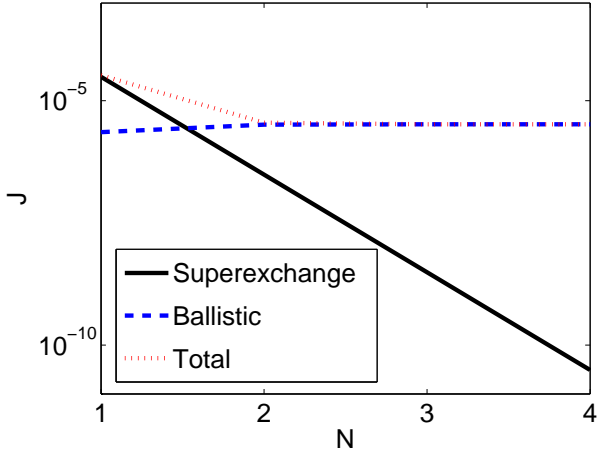}}} \caption{Energy current
as a function of bridge size assuming a superexchange (full), or a
ballistic (dashed) mechanism.
The dotted line is the total current.
Other parameters are $\epsilon=2$,
$\kappa=0.2$, $\Gamma=0.05$, $T_L=0.2$ and $T_R=0.1$.} \label{FigC}
\end{figure}


\subsection{Numerical Simulations}

The steady state dynamics of the isotropic XY model in the resonant regime is presented in
Fig. \ref{Fig3}. The left panel displays the energy current as a
function of chain size. We note that the current scales as $J\propto
N^0$, indicating on a ballistic energy transfer mechanism
\cite{Michel07}. The right panel of Fig. \ref{Fig3} presents the
behavior of the current as a function of spin gap $\epsilon$. At
high temperatures, $T_{\nu}>\epsilon$, the current follows $J\propto
\epsilon$, as expected for a ballistic motion. For large $\epsilon$,
beyond the reservoirs temperatures, the current declines since many
(high energy) subsystem modes cannot participate in the transport process
any longer as  bath modes matching the subsystem frequencies are not significantly
populated. Qualitatively, one may suggest that  $J\propto
\epsilon/(\epsilon^2+a^2)$, where $a$ is large at high temperatures.

As discussed above, the master equation method followed here cannot
be utilized once the large gaps in the band structure close, see
Fig. \ref{Fig4a}; $N=10$. Therefore, we cannot faithfully describe
here the role of the anisotropy exchange parameter $\delta$ on the
dynamics. Using a Redfield type approach without invoking the SA, it
can be shown that for large enough $\delta$, instead of the
(resonant) ballistic dynamics, heat propagates in a diffusive manner
\cite{Gemmer}. Therefore, while the off-resonance superexchange
dynamics, relaying on the bridge as a mediating medium, does not
depend on the fine details of the chain Hamiltonian, in the
resonance regime transport characteristics crucially depend on the
details of the chain structure.

\section{Conclusions}
We studied the energy transfer behavior in homogeneous linear spin
chain models coupled at the two ends to thermal reservoirs in two opposite limit:
in the off-resonance and resonance cases.
In the off-resonance limit the dynamics was investigated by adopting
the recently developed ETBO method \cite{BOheat}.
The combination of analytic manipulations and numerical simulations confirmed
that the energy current exponentially decreased with distance, an indication of a
coherent-superexchange transport mechanism. This behavior is generic,
irrespective of the details of the chain model. In the resonant
regime a standard master equation method was used, specifically demonstrating that the
energy dynamics in the isotropic XY chain model is ballistic, as the current does not
depend on the system size.

We separately presented theories for describing off-resonance and
resonance energy transmission, with the bridge modes located either
above or in resonance with the reservoirs populated modes. A
complete theory for describing, on the same footing, these two
limits could be based on a surface hopping approach \cite{Tully}, or
relaying on a nonmarkovian master equations for describing the chain
dynamics \cite{Breuer}. Here we demonstrate the crossover between
the superexchange behavior and the resonant dynamics by showing, on
the same plot, the deep-tunneling energy current, the ballistic
component, and the total current, as a function of bridge length,
see Fig. \ref{FigC}. Data was generated for the isotropic XY chain
connected to ohmic-bosonic reservoirs maintained at low
temperatures. The superexchange behavior was simulated by adopting
Eq. (\ref{eq:currFF}). The ballistic component was gained using the
method explained in Sec. IV.
We find that for short chains the coherent-superexchange contribution,
resulting from the transmission of low frequency modes across the bridge,
dominates the current. In contrast, for long chains resonant conduction is more significant,
though the population of bath modes matching the system gaps is small at low temperatures.
The turnover between the tunneling dynamics and the
resonant behavior occurs between $N=1$ to $2$ for a broad range of
parameters, $\epsilon=1-2$, $\kappa=0.05-0.2$, $\Gamma=0.01-0.05$, $T_{\nu}\sim 0.1-0.5$
(dimensionless units of energy, $\hbar\equiv 1$).
This observation lies in general agreement
with recent experiments of triplet energy transfer on $\pi$-stacked
molecules, demonstrating that the turnover between tunneling
and (resonant) diffusive mechanisms occurs between $N=$1 to 2
\cite{EETratner}.
We expect that the Heisenberg model will similarly show a turnover between
the superexchange mechanism and the diffusive (hopping) dynamics
around similar bridge sizes.

While the present analysis was mainly carried out adopting the isotropic XY
chain as the bridging object, the results of the ETBO method hold for the
anisotropic Heisenberg chain and other similar variants, as long as
gaps between different excitation manifolds are larger than energy
differences within each band. Furthermore,
the total Hamiltonian, combining the reservoirs and (nonlinear) system-bath
interactions, cannot be generally mapped onto a noninteracting
fermion model \cite{Jordan}.

The energy tunneling-superexchange behavior observed in the off-resonance
regime has been discussed before in the context of excitation energy
transfer \cite{EET}. Here it is rigorously obtained in a first
principle derivation, relaying on the timescale separation between
subsystem dynamics and the baths' motion, irrespective of the
details on the chain spectrum, the reservoir realization, and
system-baths interaction form. We expect this general behavior to
show itself in numerous systems, including organic and biological structures,
exploring electronic \cite{EETexp,EETratner} and vibrational \cite{DlottE}
energy transmission.


\begin{acknowledgments}
L.-A. Wu has been supported by the Ikerbasque Foundation Start-up,
the CQIQC grant, the Basque Government (grant IT472-10)
and the Spanish MEC (Project No.
FIS2009-12773-C02-02). DS acknowledges support from an NSERC
discovery grant.
\end{acknowledgments}




\begin{thebibliography}{9}
\bibitem{comment}
The ETBO scheme is valid for Ohmic baths, if their temperatures are
low, below the subsystem energy spacing. In such cases bath modes
overlapping with subsystem  excitations are not populated and the
dynamics is controlled by off-resonance bath modes.
\bibitem{commXY}
We refer to the chain model of Eq. (\ref{eq:Hspin}) as the "isotropic XY chain", following the
convention in the quantum information community.
In other fields, this model is sometime refered to as the XX model, to indicate that
the exchange coupling in the $x$ and $y$ directions are identical; The anisotropic Heisenberg model
(\ref{eq:HspinHeis}) is then refered to as the XXZ model, given the anisotropy in the $z$ direction.
\bibitem{comment2}
For convenience of the analytical analysis, we write the Hamitonian
such that the spin chain ground state is set to zero, rather that to
$-N\epsilon/2$ as in Eq. (\ref{eq:Hspin}).
\end{thebibliography}
\end{document}